\documentstyle[12pt,russian]{article}
\begin{document}
\begin{center}
{\bf GEOMETRISATION OF ELECTROMAGNETIC INTERACTION}
\end{center}
\begin{center}O.A.Olkhov\\
{\it Semenov Institute of Chemical Physics, Russian Academy of Sciences\\
E--mail: olega@gagarinclub.ru}
\end{center}
\medskip
\noindent     
A new concept for the geometrisation of electromagnetic interaction is proposed. 
Instead of the concept "extended field--point sources", interacting 
Maxwell's and Dirac's fields are considered as a unified closed noneuclidean
and nonriemannean space--time 4-manifold. This manifold can be considered as
geometrical realisation of the "dressed electron" idea. Within this approach, 
the Dirac equation proves to be a relation that accounts for topological and 
metric characteristics of this manifold. Dirac's spinors serve as basis vectors 
of its fundamental group representation, while the electromagnetic field 
components prove to be components of a curvature tensor of the manifold 
covering space. Energy, momentum components, mass, charge, spin and 
particle--antiparticle states appear to be geometrical characteristics of the 
above manifold.

\par\medskip
\noindent
{\bf Introduction}
\par\smallskip
First attempts of the electromagnetic field geometrisation were undertaken
just after the appearance of the theory of general relativity. The goal was 
to unify gravitation and electromagnetism within one geometrical approach
(Weyl, Kaluza, Einstein, Fok, Wheeler and others). It was expected that 
gravitation and electromagnetism can be considered as a manifistation of
noneuclidean properties of the physical space--time as it is for
gravitational field alone (see, for example, [1-3]). Later there were also
attempts of the gauge fields geometrisation where these fields were 
interpreted as connections in a space of the local gauge symmetry group
[4,5]. 

We showed early that the equation for free Dirac's field can be interpreted 
as a relation that accounts for the topological and metric properties of 
the nonorientable space--time 4--manifold which fundamental group is generated 
by four glide reflections [6--8]. (Two dimentional analog of such manifold is 
a Klein bottle [9]). We also noticed there that Maxwell's equations for a 
free electromagnetic field can be interpreted as a group--theoretic relations 
describing the orientable 4--manifold which fundamental group is generated by 
four parallel translations (two dimentional analog of such manifold is a 
torus [9]).

We shall show now that the system of equations for interacting Dirac and 
Maxwell fields can be also considered as a topological encoding of some unified 
closed connected nonorientable space--time 4--manifold.
\par\medskip
\noindent
{\bf Free electromagnetic field as an orientable space--time 4--manifold}
\par\medskip
Before the interacting fields consideration we shall firstly show that 
Maxwell's equations for free electromagnetic field can be interpreted as 
relations describing topological properties of some orientable 4--manifold. 
For more visualization let us consider the two-dimentional orientable manifold 
that is homeomorphic (topologically equivalent) to a torus. Torus can be 
represented as a product of two circles $S_1\times S_2$ [8]. Let $L_1$ and 
$L_2$  be the circles lengths. Suppose that $L_1=L_2$. Let us find out 
relations expressing topological (orientable) and metric ($L_1=L_2$) invariants 
of the manifold and let us show that such relations are formally analogous to 
Maxwell's equations. 

One of the manifold topological invariants is its fundamental group. This group 
elements are classes of pathes starting and finishing at the same point
[8]. There are two classes for our two dimentional torus and corresponding 
pathes are homeomorphic to the circles $S_1$ and $S_2$. This group is 
isomorphic (assume one--to--one correspondence) to the group of two parallel 
translations $L_1$ and $L_2$ along the Cartesian coordinates $0X$ and $0Y$ on
euclideam plane (this plane is said to be a covering surface for our torus
[8]). As the above group representation we take operators for the $L_1$--
and $L_2$--translations along $0X$ and $0Y$
$$T_x=-\frac{iL_1}{2\pi}\frac{\partial}{\partial x}, \qquad
T_y=-\frac{iL_2}{2\pi}\frac{\partial}{\partial y}.$$
It is easy to verify that the basic vectors for this representation have 
the form
$$\varphi=\exp[2\pi i(\frac{x}{L_1}+\frac{y}{L_2})].$$
Therefore, the conditions imposed by the manifold fundamental group
(parallel--translations group) and by the metric restriction ($L_1=L_2$)
can be formulated with the help of the one relation
$$\frac{\partial \varphi}{\partial x}=\frac{\partial \varphi}{\partial y}.
\eqno (1)$$

Eq.(1) does not however represent the fact that our geometrical object is 
a orientable manifold and does not therefore allow to fix the orientation. The
reason is that we choosed scalars as basic vectors for the manifold 
fundamental group representation. It is known that orientable
manifolds need more complex tensors for its representation, namely
antisymmetric second rank tensors $F_{ik}$ (bivectors)[9,11]. The bivector
components are defined by two vectors $a_i$ and $b_k$ as
$$F_{ik}=a_{i}b_{k}-a_{k}b_{i}, \eqno (2)$$
where for two-dimentional space $i=x,y; k=x,y$. 

So if we are going to change in (1) scalar for bivector we have to introduce
into the theory two vectors on the $X,Y$--plane defining the manifold 
orientation (up--down). One of the vectors is the vector of parallel 
translations ($\partial/\partial x,\partial/\partial y$). Another vector 
has to be
introduced as additional topological propriety of the manifold. Denote
this vector by ${\bf A}$. Then we have for $F_{ik}$
$$F_{ik}=\frac{\partial A_x}{\partial y}-\frac{\partial A_y}{\partial x}.
\eqno (3)$$

Let us change in (1) scalar $\varphi$ for bivector $F_{ik}$ and extend our
two-dimentional consideration to the analogous four-dimentional manifold
with the pseudoeuclidean covering space. In other words we rewrite Eqs.(1)
and (3) as 
$$\frac{\partial F_{ik}}{\partial x_{i}}=0,\quad F_{ik}=\frac{\partial A_{k}}
{\partial x_{i}}-\frac{\partial A{i}}{\partial x_{k}}, \eqno (4)$$
where $i,k=0,1,2,3;\quad x_0=ct$,\quad $c$ is a light velosity. Here (and
later on) the summation is supposed to be going over repeating indices.

We see that Eqs.(4) coincide exactly with Maxwell's equations for free field
if we consider $F_{ik}$ as the electric and magnetic fields tensor and $A_i$ as 
4--potentials [10]. This coincidence means that we can interpret the free 
electromagnetic field as the orientable closed connected space---time manifold
which fundamental group is generated by four parallel translations. The field
energy and momentum appear here as the manifold topological invariants and
the energy conservation law appears as an additional metric restriction.

We showed earlier that the equation for free Dirac's field 
can be considered as a relation describing topological and metric
characteristics of the another type manifold (nonorientable one) [6--8]. It is 
usefull now to repeat shortly the argumentation of the above 
interpretation. This equation has the form [12]:
$$\gamma^lp_l \psi=m\psi, \eqno (5)$$
where
$$\gamma^lp_l=p_0\gamma^0-p_1\gamma^1-p_2\gamma^2-p_3\gamma^3.$$
Here $m$ is a mass and $\psi$ is the four--component first rank spin--tensor. 
It can be represented by the matrix with four rows and one column
$$\psi=\left(\begin{array}{c}\xi \\ \eta \end{array}\right),\eqno (6)$$
where $\xi$ ¨ $\eta$ are two--component spinors (dotted and undotted ones).
Here $p_l=i\partial/\partial x^l$
are the 4--momentum operators, $x^0=t,\quad x^1=x,\quad x^2=y, \quad
x^3=z$, and $\gamma^l$ ($l=0,1,2,3$) are the Dirac four--row matrices. If
we choose bispinors in the form (6) then the matrices $\gamma_l$ can be 
written as
$$\gamma^0=\left(\begin{array}{cc}0&1\\1&0\end{array}\right),\quad
\gamma^{\alpha}=\left(\begin{array}{cc}0&-\sigma^{\alpha}\\ \sigma^{\alpha}
&0\end{array}\right),
\eqno (7)$$
where $\alpha = 1,2,3$ and $\sigma^{\alpha}$ are two--row Pauli matrices.
We write here four--row matrices as two--row ones: each symbol in (7)
corresponds to a two--row matrix. Here and later on $\hbar=c=1$, $\hbar$ is 
the Planck constant.

Within topological interpretation the difference between Dirac's Eq.(5)
and Maxwell's Eqs.(4) is that in (4) we have a bivector $F_{ik}$ but we have
the first--rank spin--tensor $\psi$ in (5). And we have the parallel translation 
operator $p_l$ in (4) instead of the product $p_l\gamma_l$ in (5). Any 
first--rank spin--tensor (considered as linear geometrical object) corresponds
to the geometrical structure that restores its position after rotation by
$4\pi$ (not $2\pi$) [9,11]. Such behaviour is a feature of the
nonorientable geometrical objects. (The simplest example is the M\"obius
strip [9,10])

On the other hand the $\gamma_l$ matrices can be considered within spinor basis
as a representation for the product of three symmetries with respect to
hyperplanes containing $0X$--axes [6-8]. It means that the product 
$p_l\gamma_l$ in (5) is a representation for the glide reflection group.
Therefore, Dirac's Eq.(5) can be interpreted as a metruc relation for some 
nonorientable space--time 4--manifold which fudamental group is generated by
four glide reflections and which covering space is the physical space--time
(Minkowski space). The Klein bottle is a two--dimentional analog of this
manifold [9,10].

Thus we have shown that equations for free Dirac's field and free Maxwell's
field can be interpreted as a specific mathematical description of some special
closed space--time 4--manifolds. Mass, energy and momentum components 
appear here as elements of this manifold fundamental group with dimensions 
of length. Note that the closeness of a 
manifold in pseudoeuclidean space does not imply any 
constraints on the manifold extension over the time axis. For example, a 
circle in pseudoeuclidean plane is mapped into an 
equilateral hyperbola in the usual plane [11].
In space our
manifolds are closed and bounded but they do not have a definite shape
(as any nonmetrized manifold): all manifolds obtained from some initial one
by the deformation without a damage are equivalent [9]. Nevertheless, it is
possible to indicate for these manifolds some characteristic sizes
which defined by metric conditions corresponding within geometrical 
approach to the energy and momentum
conservation laws. It is a wave length of the electromagnetic field or
the particle wave length $\hbar/p$.
\par\smallskip
\noindent
{\bf Geometrical interpretation of interacting electromagnetic and 
electron--positron fields}
\par\smallskip

Let us now consider a question of "switching on" interactions in the
geometrical representation of the above considered free fields. In other
words let us try to find out the geometrical interpretation of the
following known equations for Maxwell's and Dirac's interacting fields [12]
$$i\gamma^l(\frac{\partial}{\partial x^l}+ieA_l)\psi=m\psi,\eqno (8)$$
$$\frac{\partial F_{ik}}{\partial x^i}=j_k. \eqno (9)$$
$$F_{ik}=\frac{\partial A_i}{\partial x^k}-\frac{\partial A_k}{\partial x^i}.
\eqno (10)$$
Here $e$ is an electron charge, $m$ is an electron mass, and
the current $j_k$ is defined as
$$j_k=e\psi^+\gamma^k\psi,$$
where $\psi^+=\psi^*\gamma^0$ is so called Dirac's conjugate spinor
($\psi^*$ is a complex conjugate spinor). Here and later on we use the system
where $\hbar=c=1$.

We shall show now that Eqs.(8-10) can be considered as relations describing
topological properties of one closed 
4--manifold that has some features of both above considered ones
(corresponding to electromagnetic and electron--positron fields). This 
conclusion seems inevitable within our topological approach because it is
difficult to suggest something else. 
Up to now the topology for 4--manifolds is developed not so good as for
two-dimentional ones. For two-dimentional manifolds 
a detailed classification 
is worked out and their main topological 
invariants are defined [9,10]. Therefore, we shall try to use any possible 
parallels
between our problem and corresponding problem within two-dimentional topology.
We have in mind here that a usefulness of low-dimentional considerations
is one of the geometrical approach advatages. So let us see what could be
the result of a unification in one geometrical object proprieties of two
two-dimentional manifolds, orientable and nonorientable ones. What kind of 
object will be the hybrid of torus and the Klein bottle and how can we reflect
mathematically its topological peculiarities?

According to topological classification a two--dimentional torus is a "sphere
with one handle"  and the Klein bottle is
a "sphere with two holes covered by cross--caps or M\"obius films" [9,10].  
As a hybrid of a torus and the Klein bottle it is natural to consider a 
sphere with one handle and two cross--caps. The covering space for this 
nonorientable manifold is a hyperbolic plane and the manifold fundamental 
group is generated by glide reflections [10,14].
Let us suppose that there is some analogy between above hybrid--manifold
and the one which can represent Dirac's and Maxwell's interacting fields. Then
we can assume that Eq.(8) may be interpreted as a relation describing some
nonorientable manifold whose covering space is a four-dimentional analog of
a hyperbolic plane. Such analog is a conformal pseudoeuclidean space
(the Lobachevskian space is one of the examples [11]).

Show that Dirac's equation (8) can indeed be interpreted in this way.
Conformal euclidean space is a space that assumes conformal mapping onto
euclidean space. This means that for every point $M(x)$ of conformal euclidean
space there is a point $M_{E}$ in euclidean space where arc's differentials
are connected by the relation [11]
$$ds^2_E=f(x^0,x^1,x^2,x^3)ds^2, \eqno (11)$$
where $ds^2=g_{ik}dx^idx^k$ defines
the conformal euclidean space metrics, $ds^2_E=g^E_{ik}dx^idx^k$ is the arc's
differential squared (into our pseudoeuclidean space 
$g^E_{00}=1, g^E_{11}=g^E_{22}=g^E_{33}=-1, g^E_{ik}=0, i\ne k$).

Consider the left side of Eq.(8). As compared with Eq.(5) for
free electron--positron field it contains expression
$(\partial /\partial x^l+ieA^l)$ instead of usual derivative 
$\partial /\partial x^l$. It is customary to call this expression
"covariant derivative" because it looks like covariant derivative
$\nabla_l$  of covariant vector field $B_m$ [9-11]
$$\nabla_l B_m=\frac{\partial B_m}{\partial x^l}+\Gamma^s_{ml}B_s,
\eqno (12)$$
where $\Gamma^s_{ml}$ is a connection.

The connection geometrical meaning is that the covariant derivative plays 
the role of the parallel translation generator for the conventional tensor field
defined on some manifold (the connection for euclidean space is zero
and the parallel translation generator is a "usual" derivative 
$\partial/\partial x^l$) [9-11]. But there does not exist a connection of this
kind into arbitrary space for spintensors (in particular for 
4-component Dirac's spinors). The reason is that spintensors are the euclidean
(not affine) tensors. The transformation law for their components is defined
by the rotation group representation and it can not be extanded to the group 
of all linear transformations [13]. This means that spintensors can be compared
in two different points only if orthogonal frames remain orthogonal after 
corresponding transfer through the space.

But for particular cases---for conformal euclidean space, for example, 
we can always
map the vicinity of any point $M$ onto vicinity any another point $M'$ in
such way that the orthoganal frame at $M$ remains orthogonal in $M'$ [11].
Therefore the parallel translation for spinors in this space can be defined by
the same formulas as for any other tensors and then only the connection
components $\Gamma^p_{lp}$ will be nonzero [15].

Recall that conformal euclidean space would not be here the physical 
space --time
but the manifold covering space. This space is only a mathematical
instrument for the manifold fondamental group description. And only this
nonmetrized 4--manifold (that is not a riemann space at all) represents
interacting electromagnetic and electron--positron fields. Suppose now that
we can consider $ieA^l$ in (8) as a connection $\Gamma^p_{lp}$ in the space 
like the conformal euclidean one.(Later we shall call this space as  
"conformal pseudoeuclidean" though it is not even supposed to be riemann space). 
Then we can interpret the expression 
$(\partial/\partial x^l+ieA^l)$ in (8) as a generator of infinitesimal 
translations
in conformal pseudoeuclidean space. The parallel translation generator
multiplied by the reflection operator $\gamma^l$ gives within the spinor basis 
the local glide reflection operator [6,9,10]. It leads to the main conclusion:
Dirac's equation (8) can be interpreted as a relation                    
describing topological and metric properties of some 4--manifolds. The
manifold fundamental group is a local glide reflection and manifold's covering 
space is 
conformal pseudoeuclidean space. This manifold is infinitly connected
in contrast to four--connected manifolds describing free electron--positron and
electromagnetic fields.

Considering $ieA^l$ as a connection in the manifold covering space we can
give a geometrical interpretation for the electric and magnetic fields
components (or for components of electric and magnetic fields tensor $F_{ik}$).
Let us use for this purpose the relation betweeen the connection 
$\Gamma^k_{lm}$ and the space curvature tensor $R^q_{lk,i}$ [9,11]
$$R^q_{lk,i}=\left(\frac{\partial \Gamma^q_{li}}{\partial x^k}-
\frac{\partial \Gamma^q_{ki}}{\partial x^l}+\Gamma^q_{kp} \Gamma^p_{li}
-\Gamma^q_{lp} \Gamma^p_{ki}\right).\eqno (13)$$
(Summation is here going over repeating indices from $0$ to $3$). 

After contraction $R^q_{lk,i}$ over upper and right lower indices one obtaines
(denote the result as $R^0_{lk}$): 
$$R^0_{lk}=R^q_{lk,q}=\frac{\partial \Gamma^q_{lq}}{\partial x^k}-
\frac{\partial \Gamma^q_{kq}}{\partial x^l}.\eqno (14)$$
Comparing (14) and (10) and taking in mind that $\Gamma^q_{mq}=ieA_m$, we have
$$ieF_{ik}=R^0_{ik},$$
     Comparing Eq. (14) with Eq. (10) and using the fact that $\Gamma^q_{mq} = 
ieA_m$, one obtains
$$ieF_{ik}=R^0_{ik},\eqno (15)$$
i.e., within the geometrical interpretation, the tensor of 
electric and magnetic fields coincides, except for the factor 
$ie$, with certain components of the curvature tensor of a 
covering surface. Therefore, Maxwell's Eq. (9) relates 
the above-mentioned components of curvature tensor to the basis 
functions of the fundamental group, thereby rendering the system 
of Eqs. (8)--(10) closed. The curvature tensor for a space with 
constant curvature $K$ has the form [11]
$$R_{ij,kl}=K(g_{ik}g_{jl}-g_{il}g{jk}).\eqno (16)$$
Comparing Eqs. (16) and (15), one arrives at the conclusion that, 
within the geometrical interpretation, the electric charge $e$ is 
proportional to the covering space constant curvature $K$.

Finally, Eqs.(8-10) for interacting electromagnetic and electron--positron 
fields can be written within geometrical approach as
$$i\gamma^l(\frac{\partial}{\partial x^l}+\Gamma^p_{lp})\psi=K\psi,\eqno (17)$$
$$R^0_{ik}=\frac{\partial \Gamma^p_{ip}}{\partial x^k}-\frac{\partial
\Gamma^p_{kp}}{\partial x^i},\eqno (18)$$
$$\frac{\partial R^0_{ik}}{\partial x^i}=ie^2\psi^+\gamma^k\psi,\eqno (19)$$

\par\smallskip
\noindent
{\bf Conclusion}
\par\smallskip
Finally, we have the following geometrical 
interpretation of electromagnetic interaction.

\noindent 1. Electromagnetic field and its sources (electron--positron field) 
can be considered as a single closed infinitely connected nonorientable 
nonmetrized 4--manifold.\\
2. Covering space of this manifold is a conformal
pseudoeuclidean space.\\
3. Potentials $A_k$ is defined by the connection 
of this space $\Gamma_k$ ($ieA_k=\Gamma_k$).\\
4. Electric and magnetic field components are defined by the
components  of the covering space curvature tensor $R^0_{ik}$
($ieF_{ik}=R^0_{ik}$).\\
5. Dirac's and Maxwell's equations appear as the relations imposing metric
restictions on generators of the manifold fundamental group.\\
6. Dirac spinors appear as basic vectors for  the manifold fundamental 
group representation.\\
7. Electron charge appears as a constant covering space curvature.\\
8. Electron mass appears as a metric parameter of the manifold fundamental 
group.\\
9.Particle--antiparticle states and states with different spin projections
are the reflection of the manifold nonorientability.\\
10. In a way above manifold can be considered as "dressed electron" and it 
looks like fluctuating shapeless microscopic droplet of space--time points.\\

One comment in conclusion. Geometrisation of Dirac's equation introduces 
new topological interpretation of quantum formalism. But it is important 
that replacing a "wave--particle" by a nonmetrized 
space--time manifold does not mean "more determinism" for the quantum
object description and the topological approach
does not introduce any hidden variables and does not therefore contradict
Bell's and von Neumann's theorems [16,17].\\

\noindent
{\bf References}\\
\noindent
1. H.Weyl, Gravitation und Electrisit\"at, Berlin, Sitzber.:Preus.Akad.Wiss,
1918.\\
2. A.Einstein, Riemann Geometrie mit Aufrechterhaltung des 
Fernparallelismus,
Sitzungsber: preuss. Akad. Wiss., phys-math. K1., 1928, 217-221\\
3. J.A.Wheeler, Neutrinos, gravitation and geometry, Bologna, 1960\\
4. N.P.Konopleva, V.N.Popov, Gauge fields, Chuz--London--N.Y.: Harwood 
acad.publ., 1981.\\
5. M.Daniel, C.M.Vialett, Rev.Mod.Phys. 52(1980)175.\\
6. O.A. Olkhov, in 
Proceedings of the 7th International Symposium on Particles, Strings and
Cosmology, Lake Tahoe, California, 10-16 December 1999, p.160.\quad 
e-print quant-ph/0101137.\\
7. O.A.Olkhov, Chemical Physics (Chimitcheskay Fisika, in russian). 19, N6(2000)
13\\
8. O.A.Olkhov in Thesises of the International Seminar on 
Physics of
Electronic and Atomic Collisions, Klyasma, Moscow region, Russia, 12-16 March 
2001, p.28,\quad e-print quant-ph/0103089\\
9. B.A. Dubrovin, A.T.Fomenko, S.P.Novikov, Modern geometry---methods and
applications, N.Y.: Springer, 1990, Pt.2, Ch.4\\
10. H.S.M. Coxeter, Introduction to geometry. John Wiley and Sons, 
N-Y-London, 1961, Pt.4, Ch.21\\
11. P.K.Raschevsky, Rimanova geometria i tensorny analiz, M.:Nauka, 1967.\\
12. J.D. Bjorken, S.D. Drell, Relativistic quantum mechanics, McGraw Hill
Book Company, 1964\\
13. E. Cartan, Le\c{c}ons sur la Th\'eorie des Spineurs, 
Actualit\'es Scientifiques et Industrielles, No.643 and 701, Hermann, 
Paris, 1938\\
14. H.S.M. Coxeter, W.O. Moser,  Generators and Relations for discrete
groups. New York-London.: John Wiley and Sons, 1980.\\
15. H.Weyl, Z.f.Phys. 56(1929)330.\\
16. I.S. Bell, Rev.Mod.Phys.38(1966)447.\\
17. J.V. Neumann, Mathematische grundlagen der 
quantenmechanik, Verlag von Julius Springer, Berlin, 1932. \\

\end{document}